\documentclass[twocolumn,prl,aps,showpacs]{revtex4}

\usepackage{amssymb}
\usepackage{amsmath}
\usepackage{graphicx}
\usepackage{dcolumn}
\usepackage{bm}

\begin{document}

\preprint{031015-QCD}

\title{Nanodot-Cavity Electrodynamics and Photon Entanglement}


\author{Wang Yao}
\author{Renbao Liu}
\author{L. J. Sham}

\affiliation{Department of Physics, University of California San
Diego, La Jolla, California 92093-0319}

\date{\today}

\begin{abstract}

Quantum electrodynamics of excitons in a cavity is shown to be
relevant to quantum operations. We present a theory of an
integrable solid-state quantum controlled-phase gate for
generating entanglement of two photons using a coupled
nanodot-microcavity-fiber structure. A conditional phase shift of
$O(\pi/10)$ is calculated to be the consequence of the giant
optical nonlinearity keyed by the excitons in the cavities.
Structural design and active control, such as electromagnetic
induced transparency and pulse shaping,  optimize the quantum
efficiency of the gate operation.

\end{abstract}

 \keywords{}
\maketitle

Semiconductor nanodot plays a key role in nanoscience as has been
demonstrated by the electrical control of transport
\cite{tarucha03} and the optical control of quantum operations
\cite{li03}. Following the study of quantum electrodynamics of
atoms in cavity (CQED)  \cite{berman94}, effort is underway in the
study of CQED of excitons in nanodots
\cite{microsphere1_hailinwang}. We report here the results of a
theoretical study of excitons in CQED as illustrated by the
proposal of a solid state controlled phase gate which entangles
two photons.

Entangled photon pairs are the main stay of quantum information
processing \cite{bdsw} and the controlled gate which conditions
the dynamics of one photon on the state of the other also enables
a key logic operation for quantum computation. There are two
approaches to realize such gates: (1) linear optics with
projective measurements \cite{LOQC} and (2) nonlinear optics at
the discrete photon level. The logic gate working with few-photon
nonlinear optics requires impractical interaction length (e.g.
several meters) in conventional Kerr media \cite{Kerr_SM}. To
obtain giant optical nonlinearity for a two-photon logic gate,
novel schemes have been demonstrated, e.g. the atom-cavity QED
\cite{Phaseshift_Mabuchi_Kimble}, or proposed, e.g. slow light in
a coherently prepared atomic gas exhibiting electromagnetically
induced transparency (EIT)
\cite{Ultrslowphoton_Imamoglu}.

The relevance of excitons in CQED is strengthened by the recent
advances in solid state photonics and optoelectronics. We expect
that the localization of the optical excitations  would lead to
ready integration of the solid state cavity devices with extant
devices. Advances relevant to our proposal in semiconductor
quantum devices include single photon sources operating at room
temperature \cite{michler,lounis},
high-Q microsheres and their coupling to nanodots
\cite{microsphere1_hailinwang} and  to fibers
\cite{microsphere2_vahala}, and photonic lattice waveguides and
cavities \cite{2D_photoniclattice,fainman}.

\begin{figure}[b]
\begin{center}
\includegraphics[width=7.5cm, height=6.63cm, bb=100 320 530 700,
clip=true]{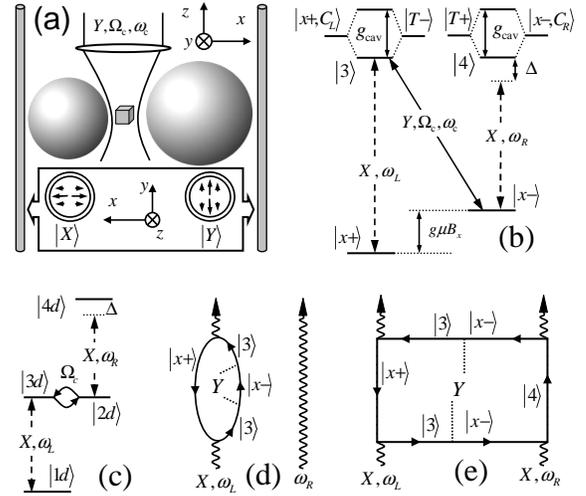}
\end{center}
\caption{The coupled system of fibers, cavities and nandot: (a)
its physical structure; (b) its energy structure, explained in
text; (c) the dressed energy states; (d) and (e) Feynman diagrams
of  the first and third order transmission processes, where the
wavy lines denote the photons, the solid lines the dressed
electron propagators, and the dotted lines connected by $Y$ the
pumping light. } \label{device1}
\end{figure}

The qubit in our scheme is represented by two polarization states
of a photon. In a quantum controlled phase gate, a two-photon
state acquires a phaseshift conditional to their polarization
configuration. The arrangement of our proposed device is given in
Fig.~\ref{device1}(a). Two photons traveling along two optical
fibers receive their interaction by coupling to two silicon
microsphere cavities which are joined by a doped nanodot. The dot
provides in theory \cite{ki3_cavity_dot} a strong third-order
optical nonlinearity which is essential for a controlled
interaction between two photons. Two cavities of different
resonant frequencies are needed to afford control of coupling to
either photon. They also act as an in-situ energy filter
preventing two photons ending in the same fiber.

The photon scattering at the phase gate have inevitably some
unwanted dynamics such as polarization-dependent reflection and
motion-polarization entanglement. By relying on the transmission
probability, the gate has the probabilistic nature as the linear
optics procedure. The essential distinction lies in our use of the
strong nonlinearity to provide definite interaction dynamics in
the cavities versus the entanglement generated by the projective
measurement. The probabilistic nature arises from the coupling of
the photons to the solid state system, which is unavoidable in any
system, but its effect can be ameliorated. Our solution is two
pronged:  to eliminate the linear reflection by EIT and to
minimize motion-polarization entanglement by pulse shaping and
system design.

The two ${\rm LP}_{11}$ modes in a step index optical fiber
\cite{singlemodefiber} 
are chosen as the two polarization states $|X\rangle$ and
$|Y\rangle$ for the qubits [see Fig. \ref{device1}(a)]. The
relevant modes in the microcavities are chosen to be the TE modes
resonant with the nanodot transitions while the other TE modes and
all TM modes are tuned far off-resonant for a small cavity ($\sim
\mu$m) \cite{Modes_microspheres}. The TE cavity mode can be
excited only by an $|X\rangle$ photon in the fiber, whose coupling
strength to the cavity on the left (right), $\kappa_{L(R)}$, are
designed by adjusting the distance between the cavity and the
fiber \cite{microsphere2_vahala}. Thus, only in the $|XX\rangle$
state do the two incoming photons  interact via the cavity-dot
coupling system, resulting a conditional phase-shift.

The strong photon-photon interaction induced by the dot-coupled
cavities is favored by both the small cavity-mode volume and the
large dipole moment of the nanodot transitions but the nanodot
which contains a single active electron plays an essential role.
The basic nonlinear optical process is illustrated with the aid of
the energy structures in Figure~\ref{device1}(b). A strong
magnetic field is applied along the $x$ direction to produce
non-degenerate transitions from the electron spin states to the
charged exciton states (trions), which are tuned respectively in
resonance with the two cavity modes. The two split electron states
are $|x\pm\rangle  \equiv (1/\sqrt{ 2})(e_{+}^{\dagger }\pm
e_{-}^{\dagger })|G\rangle $, and the two degenerate trion states
are $|T\pm\rangle \equiv (1/\sqrt{ 2})(e_{+}^{\dagger
}e_{-}^{\dagger }h_{-}^{\dagger }\pm e_{+}^{\dagger
}e_{-}^{\dagger }h_{+}^{\dagger })|G\rangle $, where
$e^\dagger_\pm$ and $h^\dagger_\pm$ create electron and hole spin
states along the $z$ axis. The transition selection rules are:
$|x\pm\rangle \leftrightarrow |T\mp\rangle $ via the $X$ polarized
field  and $|x\pm\rangle  \leftrightarrow |T\pm\rangle $ via the
$Y$ polarized field. The spatial configuration of the cavity-dot
structure is such that the TE modes are X-polarized at the site of
the nanodot. The strong coupling between the trion state
$|T{-}\rangle$ and the cavity-dot state $|x{+},C_{L}\rangle$ (or
between $|T{+}\rangle$ and the cavity-dot state
$|x{-},C_{R}\rangle$) mixes each pair into two split
trion-polariton states, where $C_{L(R)}$ denotes the left (right)
cavity mode.  We denote the lower polariton states as $|3\rangle$
and $|4\rangle$, respectively. The four states, $|x+\rangle$,
$|x-\rangle$, $ |3\rangle$, and $|4\rangle$, form the level
structure for the optical nonlinearity and all other states are
assumed far off resonance. This situation is well satisfied by the
cavity-dot coupling $g_{cav}\sim 0.5$ meV, cavity linewidth $\sim
0.1$ meV, and Zeeman splitting $g\mu _{B}B_{x}\sim 1$ meV.

To induce an interaction between the photons from the left and
right channels, a strong $Y$-polarized pump pulse is applied to
resonantly couple the states $|x-\rangle$ and $|3\rangle$.
Consider the effect of this classical field in the dressed basis:
$\left|1d\right\rangle \equiv\left| x+\right\rangle
\left|N\right\rangle$, $\left|3d\right\rangle
\equiv\left|3\right\rangle \left|N\right\rangle$,
$\left|2d\right\rangle \equiv\left|x-\right\rangle\left|
N+1\right\rangle$, and $\left|4d\right\rangle\equiv\left|
4\right\rangle \left| N+1\right\rangle,$ where the $Y$-polarized
coherent field is approximated by the Fock state $|N\rangle$ with
large $N$. Fig.~\ref{device1}(c) shows how the two $X$ photons on
separate fibers which affect separately the modes in the left and
right cavities are coupled by the $Y$ pump. The coupling strength
$\Omega_c$ between $\left|3d\right\rangle$ and
$\left|2d\right\rangle$ is proportional to the electric field
strength.  Thus, the nonlinear optical coupling  is readily
manipulated by switching on and off the pump pulse. The classical
pump pulse also increases the efficiency of the operation by
cooling the spin system and by eliminating the linear reflection
and absorption by laser cooling and EIT (see Eq.~(\ref{linearT}))
\cite{KerrEIT_Imamoglu,Ultrslowphoton_Imamoglu}.

The transformation of the polarization state of two photons is
carried out by scattering theory. The initial state is specified
by the density matrix $\rho_i$ in the basis set of the direct
products of the polarization states,  $|\sigma_L\sigma_R\rangle$
with $\sigma= X$ or $Y$, and of the wave vector states
$|k_L,k_R\rangle$.  The transmitted state $\rho_f$ is given by
$t\rho_i t^\dagger$, where $t$ is the transmission matrix. By
design, $t$ is diagonal in the polarization states. The final
density matrix of the two-photon polarizations is obtained by
tracing $\rho_f$ over the wave vectors of the photons \cite{mott}.
The transmission $t$ is obtained via the $T$ matrix conserving the
total energy. The linear and nonlinear scattering terms,
illustrated by the Feynman diagrams, Fig.~\ref{device1}(d) and (e)
indicating only the lowest order dressing terms by the $Y$ field,
are non-perturbatively calculated \cite{atomphoton} as
\begin{widetext}
\begin{subequations}
\begin{eqnarray}
&& T_{fi}^{\left( 1\right)} = \delta_{\sigma_L,X} \delta
   _{k_{R},k_{R}^{\prime }} \frac{\left|\kappa_{L}\right|^{2}}{
   2}\times \frac{E_{1d}+\hbar c k_{L}-E_{2d}}{\left(
   E_{1d}+\hbar ck_{L}-E_{3d}+i{\Gamma _{3}/2}\right) \left(
   E_{1d}+\hbar ck_{L}-E_{2d}\right) -\Omega _{c}^{2}},
  \label{linearT} \\
&& T_{fi}^{\left( 3\right) } = \frac{
    \delta_{\sigma_L,X}\kappa_{L}\Omega_c/\sqrt{2}}{\left(
    E_{1d}+\hbar ck_{L}-E_{3d}+i{\Gamma _{3}/2}\right) \left(
    E_{1d}+\hbar ck_{L}-E_{2d}\right)-\Omega _{c}^{2}}\times
   \frac{\delta_{\sigma_R,X}\left|\kappa_{R}\right|^{2}/2}{E_{1d}+\hbar
   ck_{L}^{\prime }+\hbar
   ck_{R}^{\prime }-E_{4d}+i{\Gamma_{4}/2}} \notag \\
&& \phantom{T_{fi}^{\left( 3\right) } }   \times
   \frac{\delta_{\sigma_L,X}\kappa^*_{L}\Omega^*_c/\sqrt{2}}{\left(
   E_{1d}+\hbar ck_{L}^{\prime }-E_{3d}+i{\Gamma_{3}/2}\right)
   \left( E_{1d}+\hbar ck_{L}^{\prime }-E_{2d}\right) -\Omega
   _{c}^{2}} ,
   \label{nonlinearT}
\end{eqnarray}
\end{subequations}
\end{widetext}
where  $\Gamma_{3(4)}$ is the decay rate of the polariton states
$|3\rangle$ (or $|4\rangle$).   In silicon microspheres, the
whispering gallery modes can have Q-factor as high
 as $\sim 10^8$ \cite{microsphere1_hailinwang}, so the intrinsic
decay of the cavity modes can be neglected. The relaxation rate of
trions is of the order of $\mu$eV, much less than the
cavity-to-fiber loss. Thus, the decay of the trion polaritons is
dominated by the leakage of the cavity modes into the fiber modes.
The decay rates thus can be approximated as $\Gamma_{3(4)}\approx
\left|\kappa_{L(R)}\right|^2/c$.

The $Y$-polarized photons are not scattered not being coupled to
the cavities by design. The linear term in Eq.~(\ref{linearT})
contributes to the reflection of the $X$-polarized photon from the
left channel. It is significantly suppressed by the EIT effect
\cite{KerrEIT_Imamoglu}, which results from the destructive
interference between the damped polariton state $|3d\rangle$ and
the meta-stable state $|2d\rangle$ coupled by the classical pump
field, as  is evident from the vanishing of $T_{fi}^{\left(
1\right) }$  when the incoming photon is in resonance with the
$\left|1d\right\rangle\rightarrow \left| 3d\right\rangle$
transition. The linear reflection of the $X$-polarized photons
from the right channel is eliminated since the initial state of
the system has been prepared in $|1d\rangle$ by the laser cooling
cycle: The classical pulse pumps the ground state $|x-\rangle$ to
$|3\rangle$, and then the polariton state relax to $|x+\rangle$
through the cavity-to-fiber leakage (details to be published). The
reduction of the linear reflection in both fibers brings to
prominence the third order terms which are responsible for the
gate operation. The nonlinear scattering term in
Eq.~(\ref{nonlinearT}) is composed of three fractions
corresponding to three processes: the excitation of the
trion-polariton by the left-channel photon, the induced scattering
of the right-channel photon, and the emission of the left-channel
photon by the polariton recombination.

Due to the resonance features in the T-matrix, the transmission
coefficient $t_{XX} = f  e^{-i\phi}$,  where $f$ and $\phi$  are
functions of the wavevectors, can cause the amplitude and phase
modulation of the transmitted wave since the incoming photons are
in wave packets.
 The amplitude modulation can be suppressed
either by using longer time pulses or by
working in far-resonance region. Although often overlooked in
phase-shift estimation based on $\chi^{(3)}$ susceptibility, the
phase-variation effect  results in
distortion of the pulse shape and entanglement of the motion and
polarization of the photons.
The polarization states of the two
photons are obtained by projection after the transmission. The effect
of pulse deformation may be reduced by frequency filtering the
transmitted pulse.

We show how shaping the input pulses leads to reduction of the
output pulse deformation. (1) As a consequence of the optical
coupling between the $|3d\rangle$ and $|2d\rangle$ states, the
choice of the left input photon to be within $\pm\Omega_c$ of
being in resonance with the $|1d\rangle\rightarrow |3d\rangle$
transition, the linear reflection is reduced and the first factor
on the right side of Eq.~(\ref{nonlinearT}) will yield a strong
third-order transmission. (2) To diminish the pulse distortion due
to the sharp resonant structure around the $|2d\rangle\rightarrow
|4d\rangle$ transition, the right-channel pulse is detuned about
$\Gamma_4/2$ below the transition where the real part
(corresponding to the phase-shift) of $T^{(3)}$ is large and flat
while the imaginary part (corresponding to the reflection) has
decreased to a small value.  (3) To minimize the pulse broadening
and distortion from the convolution of the input pulses with
energy conservation, we choose the two input pulses to have
sqaure-shaped spectra with much different widths. In our design,
$\Gamma_4/2$ is much larger than $\Omega_c$, so the right-channel
pulse is set the wider in frequency.

\begin{figure}[t]
\begin{center}
\includegraphics[width=7cm, height=3cm, bb=10 330 595 580, clip=true]{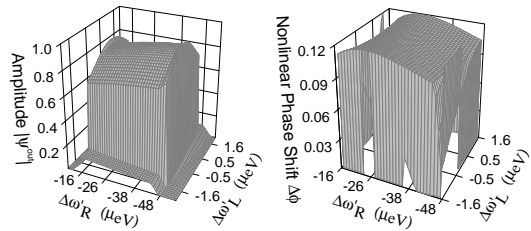}
\end{center}
\caption{The amplitude and phase of the transmitted two-photon
wavefunction as functions of the detuning
$\Delta\protect\omega_{L}\equiv \hbar ck'_L-(E_{3d}-E_{1d})$ and
$\Delta\protect\omega_{R}\equiv\hbar ck'_R-(E_{4d}-E_{2d})$. The
parameters are: $\Gamma_{3}=\Gamma_{4}=60$ $\mu$eV, $\Omega_c=8$
$\mu$eV; $g_{cav}=0.5$ meV. The input wavefunction is such that
$\Psi^{i}(k_L,k_R)=\theta(48+\Delta\omega_R)
\theta(-16-\Delta\omega_R) \theta(1.6+\Delta\omega_L)
\theta(1.6-\Delta\omega_R)$ with arguments in units of $\mu$eV.
}\label{12}
\end{figure}

\begin{figure}[b]
\includegraphics[width=7.16cm, height=3cm, bb=10 410 595 655,
clip=true]{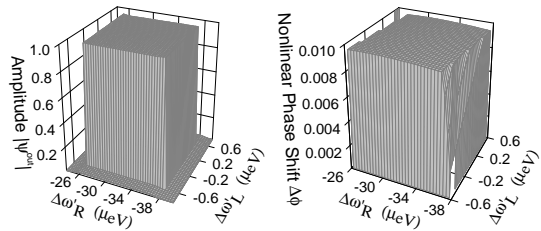} \caption{The same as Fig.~\ref{12} except that
the parameters are: $\Gamma_3=45$ $\mu$eV, $\Gamma_4=60$ $\mu$eV;
$\Omega_c=15$ $\mu$eV; $g_{cav}=0.5$ meV and $\Psi^{i}(k_L,k_R)=
\theta(38+\Delta\omega_R)\theta(-26-\Delta\omega_R)
\theta(0.6+\Delta\omega_L)\theta(0.6-\Delta\omega_R)$.}\label{01}
\end{figure}


Figure~\ref{12} presents the transmitted wavefunction ($k'_{L}$,
$k'_{R}>0$) for incoming photons in square pulses with the
polarization state $|XX\rangle$. Though visible, the pulse
distortion and broadening and the inhomogeneity in phase-shift is
quite small. A conditional phase shift of ${\pi/29}$ is obtained
with a transmission probability of $0.72$ and a 0.99 fidelity.

The loss in transmission and the pulse distortion in Fig.~\ref{12}
result mainly from the imperfect EIT when the photon is
off-resonant with the $|1d\rangle\rightarrow |3d\rangle$
transition. Improvement of both the pulse shape and transmission is effected
by increasing the pump power $\Omega_c$ (in order to open a larger
EIT window) and by using narrower bandwidth pulses at the expense of
a weak nonlinear phase shift.  An example is shown in Fig.~\ref{01},
in which a nonlinear phase shift $\sim{\pi/330}$ is
obtained almost without pulse-shape change or reflection loss, shown
by the computed
transmission probability of $\sim 0.982$ and almost perfect fidelity.

Although a small entanglement suffices for some quantum
information purposes \cite{Cirac_small_entanglement}, the small
phase-shift is not useful for most two-qubit operations. A large
phase-shift can be achieved by using a series of many identical
gates integrated into a single chip. With modern fabricating
techniques, the integrated quantum gates can be constructed either
with micro-disks and wave guides etched on semiconductor
heterostructures \cite{waveguide_cavity} or with point- and
line-defects engineered in photonic lattices
\cite{2D_photoniclattice}.

To use the system to produce an entangled photon pair rather than
to perform a controlled phase operation, we optimize the
entanglement by a different procedure. The quantum operation is
favored by maximizing the transmission but the entanglement is
favored by symmetrizing the two photons for maximal projection of
the polarization degrees of freedom. First the input state is
prepared as the equal linear combination of the four polarization
states of the two photons. Then the state is passed $n$ times
through the coupled system as described above, undergoes single
bit operation swapping the $|X\rangle$ and $|Y\rangle$ states in
both photons, and is passed through the phase gate $n$ more times.
This symmetrizes the transmitted density matrix after projecting
out the motional degrees of freedom. Table~\ref{table2} shows the
calculated results for 2+2 and 4+4 gates. The transmission
probabilities $T$ are much lower than for the phase operation. The
quantitative measures of the operation including fidelity ${\rm
Tr}\left[ \rho _{t}\rho _{ideal}\right]$  towards the maximally
entangled state $\frac{1}{\sqrt{2}}(|XY\rangle +|YX\rangle)$, the
purity ${\rm Tr}\left[ \rho _{t}^{2}\right]$, the concurrence $C$
and the entanglement of formation $E(C)$ \cite{wootters}, all show
excellent entanglement.

\begin{table}[b]
\caption{n+n gates with parameters:
$\protect\gamma_{ch}^L=0.12meV; \protect \gamma_{ch}^R=1meV;
\Omega_c=6.2\protect\mu eV$. The left- and right-channel Gaussian
pulses with FWHM $7.5\protect\mu eV$ and $50 \protect \mu eV$ are
resonant with the left and right polariton transitions,
respectively. $C$ denotes the concurrence and $E{(C)}$ the
entanglement of formation. } \label{table2}
\begin{ruledtabular}
\begin{tabular}{cccccc}
n & Transmission & Fidelity & Purity & C & $E(C)$  \\
2 & 0.1165 & 0.8638& 0.9591 & 0.7277 & 0.6272  \\
4 & 0.02074 & 0.9758 &0.9850 & 0.9515 & 0.9306 \\
\end{tabular}
\end{ruledtabular}
\end{table}

In summary, we have proposed a solid-state controlled phase gate
for two photons. The flying qubits are conducted through fibers
coupled to scattering centers composed of microcavities connected
by a doped semiconductor nanodot. This allows a fiber
implementation of quantum information processor. Calculated
results show that the system is flexible as a phase gate as well
as producing strong entanglement. The trions in doped nanodot used
for nonlinear interaction here can be replaced by other electronic
systems, such as biexcitons in an undoped nanodot (results will be
published elsewhere), states in nanoclusters, or even some strong
transitions in rare-earth impurities, e.g., the 4d-5f transition
in $Er^{2+}$. The microcavity may be microspheres or defects in
photonic lattices. The structure has unique features, such as
small size, integrability, and stability, useful for quantum
information and for scalable quantum computing.

This Work was supported by  NSF DMR-0099572, ARDA/ARO
DAAD19-02-1-0183, and QuIST/AFOSR F49620-01-1-0497. LJS thanks Y.
Fainman for helpful discussions.

\end{document}